\documentclass[11pt,a4paper]{article}
\pdfoutput=1
\usepackage{jheppub}
\usepackage{amsthm}
\usepackage[page]{appendix}
\numberwithin{equation}{section}
\numberwithin{figure}{section}
\usepackage{tikz}
\usetikzlibrary{calc}

\title{Superconformal quantum mechanics and the exterior algebra}
\author{Andrew Singleton}
\affiliation{Department of Applied Mathematics and Theoretical Physics, University of Cambridge, \\
Cambridge, CB3 0WA, UK}
\emailAdd{A.Singleton@damtp.cam.ac.uk}
\abstract{We extend the differential form representation of $\mathcal{N} = (n,n)$ supersymmetric quantum mechanics to the superconformal case. We identify the superalgebras occurring for $n = 1,2,4$, give necessary and sufficient conditions for their existence, and give explicit geometric constructions of their generators and commutation relations. Quantum mechanics on the moduli space of instantons is considered as an example.}
\arxivnumber{1403.4933}

\begin{document}
\maketitle

\section{Introduction} \label{section:intro}
Quantum mechanical systems with superconformal symmetry arise in a variety of contexts in field theory, typically as special limits or truncations of superconformal field theories. For instance, the low energy moduli space dynamics of BPS solitons is a supersymmetric mechanical model which may also be conformally invariant if the underlying field theory is superconformal. A classic example, which we consider in section \ref{subsection:example}, is quantum mechanics on the moduli space $\mathcal{M}_{k,N}$ of $k$ $SU(N)$ Yang-Mills instantons on $\mathbb{R}^4$. This arises in the discrete light cone quantisation of the $(2,0)$ theory on $N$ M5-branes \cite{Aharony:Berkooz:Seiberg:1997} and its superalgebra has been understood from this point of view. We give an alternative construction via the hyper-K\"{a}hler quotient \cite{Hitchinetal:1987} which can be understood as a sufficient condition to obtain one $\mathcal{N}=(4,4)$ superconformal model from another. Superconformal mechanical models also have a surprising connection to the near-horizon geometry of extremal black holes, tying in with the $AdS_2/CFT_1$ correspondence \cite{Gibbons:Townsend:1998,Michelson:Strominger:1999a,Clausetal:1998,BrittoPacumio:Michelson:Strominger:Volovich:1999}.

It has been known for some time that the structure of supersymmetric quantum mechanics is closely tied to the geometry of the target manifold $M$ \cite{AlvarezGaume:Freedman:1981,Witten:1982b,Witten:1982a,Coles:Papadopoulos:1990}, and it is this formalism that we exploit in constructing our superconformal algebras. In particular, we work almost exclusively in the language of wavefunctions, which in the supersymmetric context means the complex differential forms $\Omega^*(M;\mathbb{C})$. Alternative approaches include the constructions of \cite{Kuznetsova:Toppan:2011,Khodaee:Toppan:2012} using worldline differential operators, and a variety of superspace formulations reviewed in \cite{Fedoruk:Ivanov:Lechtenfeld:2011}. Our construction naturally extends the work of \cite{FOF:Kohl:Spence:1997}, which constructs the supersymmetry algebra and R-symmetry generators of the models we are interested in and explains the link to (hyper-)K\"{a}hler geometry. The necessary and sufficient conditions we derive are in parallel with those of \cite{Michelson:Strominger:1999}, which considers a similar problem in the case of chiral $\mathcal{N}=(0,n)$ superconformal mechanics, and many of our initial ans\"{a}tze for the forms of operators are inspired by their work. It's worth noting that remarkably similar structures to those we derive have appeared in the context of higher spin superparticles and the worldline formulation of higher form gauge fields \cite{Hallowell:Waldron:2007,Burkart:Waldron:2008}.

In section \ref{section:SUSY} we review the relation of SUSY to the differential form language of \cite{Witten:1982b,Witten:1982a} and the conditions for existence of extended supersymmetry \cite{AlvarezGaume:Freedman:1981,FOF:Kohl:Spence:1997}. In section \ref{section:SCQM} we derive our main results and consider the example of instanton quantum mechanics. In each case, the explicit form of all generators and their commutation relations are given in an appendix.

In summary, we find that:
\begin{itemize}
\item $\mathcal{N}=(1,1)$ supersymmetry extends to $\mathfrak{su}(1,1|1)$ superconformal invariance if and only if the Riemannian manifold $M$ admits a homothety $\mathcal{L}_Dg = 2g$ which is closed in the sense that $D_{\mu}=\partial_{\mu}K$ for some function $K$ satisfying $K=\frac{1}{2}g^{\mu \nu}\partial_{\mu}K\partial_{\nu}K$.
\item $\mathcal{N}=(2,2)$ SUSY on a K\"{a}hler manifold extends to $\mathfrak{u}(1,1|2)$ if and only if the homothety is also holomorphic, so $\mathcal{L}_DI = 0$, and the function $K$ is the K\"{a}hler potential.\footnote{The K\"{a}hler potential is not really a function, rather a section of a line bundle. This issue is discussed in section \ref{subsection:Kaehlercase}.}
\item $\mathcal{N}=(4,4)$ SUSY on a hyper-K\"{a}hler manifold extends to a real form of $\mathfrak{osp}(4|4)$ if and only if the conditions of the K\"{a}hler case are met with respect to each complex structure. In particular $K$ must be a hyper-K\"{a}hler potential, which we show to always exist in our models.
\end{itemize}
These results for $\mathcal{N} = (1,1)$ and $(4,4)$ are in line with expectations from the established literature on superconformal mechanics (see \cite{Fedoruk:Ivanov:Lechtenfeld:2011} and references therein for a fairly recent review). In particular, the superalgebras obtained fit into the classifications of \cite{Nahm:1977,Ivanov:Lechtenfeld:Sutulin:2007}. In the case of 4 supercharges, there are a number of models in the literature \cite{Wyllard:1999,Krivonos:Lechtenfeld:2010,Ivanov:Krivonos:Lechtenfeld:2003,Michelson:Strominger:1999} which admit the larger $D(2,1;\alpha)$ invariance. It should be emphasised that the fermionic content of our model is subtly different to these cases (being of type $(2,2)$ rather than $(0,4)$) and that there is no conflict in the fact that the symmetry appears not to extend to $D(2,1;\alpha)$. While we do not claim to make a definitive no-go statement in this direction, we comment on a possible obstruction in section \ref{subsection:supercharges}.

\section{Supersymmetric quantum mechanics} \label{section:SUSY}
In this section we review the representation of supersymmetric QM in terms of the exterior algebra $\Omega^*(M;\mathbb{C})$. Our basic model is that of \cite{AlvarezGaume:Freedman:1981}, namely the $\mathcal{N}=(1,1)$ supersymmetric extension of geodesic motion on a Riemannian manifold $(M,g)$:
\begin{equation} S = \frac{1}{2}\int dt \; g_{\mu \nu}\dot{X}^{\mu}\dot{X}^{\nu} + i g_{\mu \nu} \bar{\psi}^{\mu}\gamma^0 \frac{D}{Dt}\psi^{\nu} + \frac{1}{6}R_{\mu \nu \rho \sigma} \left(\bar{\psi}^{\mu}\psi^{\rho}\right) \left(\bar{\psi^{\nu}}\psi^{\sigma}\right) \label{eqn:action} \end{equation}
Here $X^{\mu}$ are coordinates on $M$ and $\psi^{\mu}$ are their 2-component Majorana fermionic superpartners, to be understood as Grassmann-odd sections of the complexified tangent bundle $T(M;\mathbb{C})$. $D/Dt$ is the appropriate covariant derivative
\[ \frac{D}{Dt}\psi^{\mu} = \nabla_{\dot{X}}\psi^{\mu} = \dot{\psi}^{\mu}+\dot{X}^{\nu}\Gamma^{\mu}_{\nu \rho}\psi^{\rho} \]
and we work in a basis where
\[\psi^{\mu A} = \frac{1}{\sqrt{2}}\left(\begin{array}{c} \psi^{\mu} \\ \psi^{\mu \dagger} \end{array} \right), \quad \gamma^0 = -\sigma^3, \quad \bar{\psi} = \psi^{\dagger}\gamma^0.\]
The model is by construction invariant under the $\mathcal{N}=(1,1)$ supersymmetry
\begin{subequations} \begin{align} \delta X^{\mu} &= \bar{\epsilon} \psi^{\mu} \\ \delta \psi^{\mu A} &= -i\dot{X}^{\mu} (\gamma^0 \epsilon)^A - \Gamma^{\mu}_{\nu \rho} (\bar{\epsilon}\psi^{\nu})\psi^{\rho A} \end{align} \end{subequations}
with corresponding Noether supercharge
\begin{equation} Q^A = g_{\mu \nu} \dot{X}^{\mu} \psi^{\nu A}. \label{eqn:supercharge} \end{equation}

Alvarez-Gaum\'{e} and Freedman have classified the manifolds for which this model admits additional supersymmetries \cite{AlvarezGaume:Freedman:1981}. They find an extension to $\mathcal{N}=(2,2)$ SUSY if and only if $M$ is K\"{a}hler, so has a covariantly constant complex structure $I$ such that the metric $g$ is Hermitian and the K\"{a}hler form $\omega_{\mu \nu} = I^{~\rho}_{\mu}g_{\rho \nu}$ is closed. $\mathcal{N}=(4,4)$ follows by demanding that $M$ is hyper-K\"{a}hler, that is $M$ has a triplet of complex structures $I^a$ satisfying the quaternion algebra
\[ I^a I^b = -\delta^{ab} + \epsilon^{abc}I^c \]
and is K\"{a}hler with respect to all three. Furthermore, \cite{AlvarezGaume:Freedman:1981} shows that these additional supersymmetries may be obtained via the replacement $\psi^{\mu A} \mapsto I_{\nu}^{a  \mu} \psi^{\nu A}$, so that the additional supercharges take the schematic form
\begin{equation} Q^{aA} \sim g_{\mu \nu} \dot{X}^{\mu} I_{\rho}^{a  \nu} \psi^{\rho A}. \label{eqn:supercharges} \end{equation}

Witten \cite{Witten:1982b} demonstrated that the canonical quantisation of (\ref{eqn:action}) leads to the exterior algebra on $M$. More precisely, the canonical commutation relations for the fermions are
\begin{equation} \left\{ \psi^{\mu},\psi^{\nu}\right\} = 0, \quad \left\{\psi^{\mu},\psi^{\nu \dagger}\right\} = g^{\mu \nu},\label{eqn:ACRs} \end{equation}
while the bosons obey the standard $\left[X^{\mu},P_{\nu}\right] = i\delta^{\mu}_{\nu}$. Note that the canonical bosonic momentum $P_{\mu}$ does not commute with a fermion carrying a curved index\footnote{For the purposes of canonical quantisation, one should work with a vielbein basis $\psi^{\alpha} = e^{\alpha}_{\mu}\psi^{\mu}$ for the fermions.} $\mu$, and a more natural choice is the covariant momentum 
\[ \Pi_{\mu} = g_{\mu \nu}\dot{X}^{\nu} = P_{\mu} -\tfrac{i}{2}g_{\nu \rho}\bar{\psi}^{\nu}\gamma^0\Gamma^{\rho}_{\mu \sigma}\psi^{\sigma}\]
defined in \cite{Michelson:Strominger:1999}, which satisfies the commutation relations
\[ \left[X^{\mu},\Pi_{\nu}\right] = i\delta^{\mu}_{\nu},\quad \left[\Pi_\mu,\psi^{\nu}\right] = i\Gamma^{\nu}_{\mu \rho}\psi^{\rho}.\]

Canonical quantisation now proceeds as follows. In the purely bosonic case, the Hilbert space $\mathcal{H}$ is identified with square-integrable complex wavefunctions $\left\langle X|f\right\rangle = f(X)$ on $M$ with the standard $L^2$ inner product
\[ \left\langle \alpha |\beta \right\rangle = \int_M \sqrt{g} \,\alpha \bar{\beta}.\]
To extend to fermions, identify $\psi^{\mu}$ as annihilation operators and $\psi^{\mu \dagger}$ as creation operators and build a Fock space in the usual way. By virtue of the commutation relations (\ref{eqn:ACRs}), the state
\[ \left| \alpha \right\rangle = \frac{1}{r!}\psi^{\mu_1 \dagger} \dots \psi^{\mu _r \dagger} \left| \alpha_{\mu_1 \dots \mu_r} \right\rangle\]
can be identified with the r-form $\alpha$ with components $\alpha_{\mu_1 \dots \mu_r}$ and the natural norm is simply the $L^2$ inner product
\[ \left\langle \alpha | \beta \right\rangle = \int_M \alpha \wedge *\bar{\beta}.\]
When $M$ is not compact, we will typically need a stronger constraint on the decay of states in $\mathcal{H}$ at infinity than just $L^2$-normalisability to ensure that everything we do is well-defined. For the most part we leave this implicit, but exponential decay will certainly suffice.

We arrive at the following dictionary between objects in QM and differential geometry:
\begin{equation} \begin{array}{c|c} \mbox{ QM } & \mbox{ DG } \\ \hline \mathcal{H} & \Omega^*(M;\mathbb{C}) \\ \left| \alpha \right\rangle \in \mathcal{H} & \alpha \in \Omega^* \\ \psi^{\mu \dagger} & dX^{\mu}\wedge \\ \psi^{\mu} & g^{\mu \nu}i_{\partial_\nu} \\ P_{\mu} & -i\partial_{\mu} \\ \Pi_{\mu} & -i\nabla_{\mu}  \end{array} \label{eqn:canonicaldictionary} \end{equation}
where $i_V$ is the contraction of a vector field $V$ with a form. Notice in particular that these operators have adjoints
\[ \psi^{\mu \dagger} = \left(\psi^{\mu}\right)^{\dagger}, \quad \Pi_\mu^{\dagger} = \Pi_{\mu},\quad f(X)^{\dagger} = \bar{f}(X).\]
\subsection{Construction of the superalgebra} \label{subsection:supercharges}
It is shown in \cite{Witten:1982b,Witten:1982a} that this model gives a particularly simple realisation of the supersymmetry algebra. With the above identifications\footnote{Of course we make a convenient choice of operator ordering here.} we see that the supercharges (\ref{eqn:supercharge}) become
\begin{subequations} \label{eqn:SUSYd} \begin{align} Q & = i\psi^{\mu \dagger} \Pi_{\mu} ~\mapsto  d \\  Q^{\dagger} & = -i \Pi_{\mu} \psi^{\mu} \mapsto  d^{\dagger} \end{align} \end{subequations}
where $d^{\dagger}=(-1)^{mr+m+1}*d*$ is the usual coderivative and $m = \mbox{dim}_{\mathbb{R}}M$. The supersymmetry algebra then gives
\[ \left\{Q,Q^{\dagger}\right\} = 2H \quad \Rightarrow  \quad H = \tfrac{1}{2}\Delta\]
where $\Delta = dd^{\dagger}+d^{\dagger}d$ is the Hodge Laplacian.

In the case of extended supersymmetry, the form of the supercharges (\ref{eqn:supercharges}) suggests an identification with other derivatives on $M$ such as the Dolbeault operator $\partial$. The exact relationship is derived in \cite{FOF:Kohl:Spence:1997}, where the requirement of extended SUSY is shown to be equivalent to the K\"{a}hler identities. Recall that these read:
\begin{subequations} \label{eqn:Kaehleridentities} \begin{align}
\left[\omega \wedge, \partial\right] = \left[\omega \wedge, \bar{\partial}\right] = 0, \quad & \left[\omega \wedge,\partial^{\dagger}\right] = i\bar{\partial}, \quad \left[\omega \wedge, \bar{\partial}^{\dagger}\right] = -i\partial \\ \left[\omega \wedge, (\omega \wedge)^{\dagger}\right] &= 2(r-n) \end{align}\end{subequations}
where $n=\mbox{dim}_{\mathbb{C}}M$. Since the construction will be important for the superconformal case, we give some detail now.

Verbitsky has demonstrated \cite{Verbitsky:1990} the existence of an $\mathfrak{so}(5)$ action on the cohomology of hyper-K\"{a}hler manifolds, extending the perhaps more familiar $\mathfrak{so}(3)$ action, known to mathematicians as the Lefschetz decomposition, existing in the K\"{a}hler case. Through the usual Hodge identification of cohomology with harmonic forms, and hence \cite{Witten:1982a} with supersymmetric ground states, it is apparent that these actions will simply be the R-symmetry groups of our model. Indeed, \cite{FOF:Kohl:Spence:1997} shows that the K\"{a}hler and hyper-K\"{a}hler cases follow respectively from dimensional reduction of minimal supersymmetry in 4 and 6 dimensions.

In the K\"{a}hler case we have a $U(1)$ action of the complex structure on r-form states, namely
\begin{equation} \left(U(\theta)\alpha\right)(V_1,\dots,V_r) = \alpha(U(\theta)V_1,\dots,U(\theta)V_r) \label{eqn:Cstructureaction} \end{equation}
where we defined
\[ U(\theta) = \exp{(iR\theta)}, \quad U(\pi) = I\]
so that this is just the extension of the action $\alpha_{\mu} \mapsto I_{\mu}^{~\nu}\alpha_{\nu}$ on 1-forms to the tensor product. It follows that $R$ acts on $(p,q)$ forms as
\begin{equation} \left. R\right|_{\Omega^{p,q}} = \tfrac{1}{2}\left(p-q\right) ~ \Rightarrow ~ R = -\tfrac{i}{2}\psi^{\mu \dagger}I_{\mu}^{~\nu}\psi_{\nu} \label{eqn:RI}\end{equation}
and in particular is hermitian, so that the group action is unitary. Using this action, we construct new supercharges
\begin{subequations} \label{eqn:Iconjugation} \begin{align} \tilde{Q} &= I^{-1}QI = i(\bar{\partial}-\partial) := d^I \\ \tilde{Q}^{\dagger} &= I^{-1}Q^{\dagger}I = i(\partial^{\dagger}-\bar{\partial}^{\dagger}). \end{align} \end{subequations}
Standard facts about K\"{a}hler geometry, in particular $\Delta_{\partial}=\Delta_{\bar{\partial}}=\frac{1}{2} \Delta_d$ along with the K\"{a}hler identities (\ref{eqn:Kaehleridentities}) then imply that the only non-vanishing commutators are
\begin{subequations} \begin{align} \left\{Q,Q^{\dagger}\right\} &= \left\{\tilde{Q},\tilde{Q}^{\dagger}\right\} = 2H \\ \left[R,Q\right] &= \tfrac{i}{2}\tilde{Q}, \quad \left[R,\tilde{Q}\right] = -\tfrac{i}{2}Q. \end{align} \end{subequations}
In particular, we get our $\mathcal{N}=(2,2)$ supersymmetry algebra and $R$ generates a $U(1)$ R-symmetry with half-integer charges.

To construct the $SU(2)$ R-symmetry we further exploit the K\"{a}hler identities. Defining the operators
\begin{subequations} \label{eqn:Kaehlersu2} \begin{align} J_+ & = \omega \wedge = \tfrac{1}{2} \omega_{\mu \nu} \psi^{\mu \dagger}\psi^{\nu \dagger} \label{eqn:J+}\\ J_- & = (J_+)^{\dagger} = \tfrac{1}{2} \omega_{\mu \nu}\psi^{\nu}\psi^{\mu} \label{eqn:J-} \\ J_3 & = \tfrac{1}{2}(r-n) = \tfrac{1}{2}(\psi^{\mu \dagger}\psi_{\mu} -n) \label{eqn:J3} \end{align}\end{subequations}
we immediately get the $\mathfrak{su}(2)$ algebra as in \cite{FOF:Kohl:Spence:1997}
\[ \left[J_+,J_-\right] = 2J_3, \quad \left[J_3,J_{\pm}\right]=\pm J_{\pm}. \]
Furthermore, again by virtue of the K\"{a}hler identities, $\mathfrak{su}(2)$ commutes with $H$ and the pairs $(Q,\tilde{Q}^{\dagger})$ and $(\tilde{Q},Q^{\dagger})$ transform as doublets. Finally, since the K\"{a}hler form $\omega$ is of type (1,1), the $SU(2)$ and $U(1)$ actions commute. Putting everything together, we see that a K\"{a}hler manifold admits $\mathcal{N}=(2,2)$ supersymmetric quantum mechanics with an $SU(2) \times U(1)$ R-symmetry acting purely on the fermions. Notice that $J_3$ exists and generates a $U(1)$ R-symmetry even in the Riemannian $\mathcal{N}=(1,1)$ case.

We mentioned in section \ref{section:intro} that a number of superconformal mechanical models with four supercharges admit $D(2,1;\alpha)$ invariance, which enjoys the larger $SU(2) \times SU(2)$ R-symmetry. This appears to be strictly larger than our model admits, and the discrepancy should arise from the subtle difference between $\mathcal{N}=(2,2)$ and $(0,4)$ supersymmetry in quantum mechanics. In $(0,4)$ the fermions form real representations of an $SO(4)$ R-symmetry, so this is certainly compatible with $D(2,1;\alpha)$ (see \cite{Michelson:Strominger:1999} for a geometric construction). However, since the representation is real, there is no canonical splitting between creation and annihilation operators in the Fock space quantisation (\ref{eqn:ACRs}) and it is therefore difficult to see how an exterior algebra representation can be built in general. Indeed, choosing such a splitting amounts to picking a preferred $SU(2)$ from which to take the Cartan generator $J_3$, which breaks $SU(2) \times SU(2) \rightarrow SU(2) \times U(1)$. While this makes an exterior algebra realisation of $D(2,1;\alpha)$ seem unlikely, we do not absolutely rule out a more subtle construction. Note that the resulting $SU(2) \times U(1)$ R-symmetry is exactly the one which appears naturally in $\mathcal{N}=(2,2)$ supersymmetry.

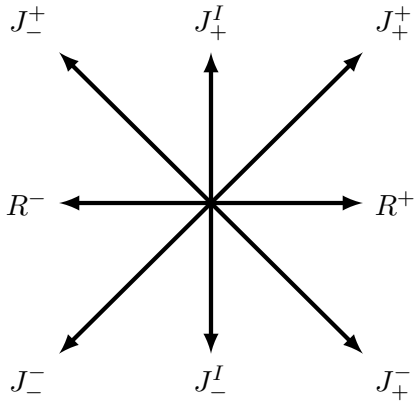
\begin{figure}
\begin{center}
\begin{tikzpicture}
\coordinate (Cartan) at (0,0);
\coordinate (Rminus) at (-2,0);
\coordinate (Rplus) at (2,0);
\coordinate (JIminus) at (0,-2);
\coordinate (JIplus) at (0,2);
\coordinate (Jplusplus) at (2,2);
\coordinate (Jminusplus) at (2,-2);
\coordinate (Jplusminus) at (-2,2);
\coordinate (Jminusminus) at (-2,-2);
\draw [ultra thick, -latex, black] (Cartan) -- (Rminus)
  node [left] {$R^-$};
\draw [ultra thick, -latex, black] (Cartan) -- (Rplus)
  node [right] {$R^+$};
\draw [ultra thick, -latex, black] (Cartan) -- (JIplus)
  node [above] {$J^I_+$};
\draw [ultra thick, -latex, black] (Cartan) -- (JIminus)
  node [below] {$J^I_-$};
\draw [ultra thick, -latex, black] (Cartan) -- (Jplusplus)
  node [above right] {$J^+_+$};
\draw [ultra thick, -latex, black] (Cartan) -- (Jplusminus)
  node [above left] {$J^+_-$};
\draw [ultra thick, -latex, black] (Cartan) -- (Jminusplus)
  node [below right] {$J^-_+$};
\draw [ultra thick, -latex, black] (Cartan) -- (Jminusminus)
  node [below left] {$J^-_-$};
\end{tikzpicture}
\caption{The root lattice of $\mathfrak{usp}(4)$ in our basis.}
\end{center}
\label{fig:rootlattice}
\end{figure}

The hyper-K\"{a}hler case now essentially follows from the above \cite{FOF:Kohl:Spence:1997,Verbitsky:1990}. For each complex structure $I^a$ we have an $\mathfrak{su}(2)_a$ and a $\mathfrak{u}(1)_a$, giving a total of 8 real supercharges
\[ Q,Q^{\dagger},Q^a,Q^{a \dagger}\]
satisfying the $\mathcal{N}=(4,4)$ supersymmetry algebra. The new ingredient is the occurrence of cross terms between the $\mathfrak{su}(2)_a$ and $\mathfrak{u}(1)_b$ parts of the algebra. In fact, a simple component computation using the definitions (\ref{eqn:RI}, \ref{eqn:Kaehlersu2}) gives the following commutators (notice that each $\mathfrak{su}(2)_a$ has the same $J_3$):
\begin{subequations} \label{eqn:SUSYso5} \begin{align} \left[J^a_+,J^b_-\right] &= 2\left(\delta^{ab}J_3 - i\epsilon^{abc}R^c\right) \\ \left[J^a_{\pm},R^b \right] &= -i\epsilon^{abc}J^c_{\pm} \\ \left[R^a,R^b\right] &= -i\epsilon^{abc}R^c.\end{align} \end{subequations}
It is simple to check that this does indeed give an $\mathfrak{so}(5)$ R-symmetry. For instance, we can choose a basis of roots by letting $J_3$ and $R^I$ span a Cartan subalgebra and defining
\[ R^{\pm} = R^J \pm iR^K, \quad J_{\pm}^{\pm} = J_{\pm}^J \pm iJ_{\pm}^K.\]
It follows from the commutation relations (\ref{eqn:SUSYso5}) that the root lattice of $\mathfrak{usp}(4)$ may be identified with figure \ref{fig:rootlattice}.
The supercharges may be collected into a single object $Q^i$ transforming in the $\mathbf{4}$ of $\mathfrak{usp}(4)$, so that the commutation relations read
\[ \left\{Q^i,\bar{Q}_j\right\} = 2\delta^i_jH.\]

\section{Extension to superconformal invariance} \label{section:SCQM}
In extending the supersymmetric systems of section \ref{section:SUSY} to superconformal invariance, we will really see the power of the geometric formalism. Michelson and Strominger have considered a similar problem \cite{Michelson:Strominger:1999} in the case of $(\mathcal{N},0)$ superconformal invariance and our results are largely parallel to theirs, though the method of derivation is rather different.

\subsection{Conformal invariance} \label{subsection:conformal}
We start by looking for conditions for an $\mathfrak{so}(2,1)$ conformal algebra\footnote{We use the same notation for an operator and its corresponding geometric object whenever this does not cause confusion. Hats will be added to operators where necessary.} $(D,H,K)$ represented by hermitian operators on r-forms. We make the assumption that $H = \frac{1}{2} \Delta$ as previously, and since under $D$ we have
\[ X^{\mu} \mapsto X^{\mu} + \epsilon D^{\mu} \]
we insist that $D$ corresponds to the flow of some vector field $D = D^{\mu}(X)\partial_{\mu}$, that is to a Lie derivative $\mathcal{L}_D$. More precisely, since this is not self-adjoint we make the ansatz
\[ D = \frac{i}{2}\left(\mathcal{L}_D^{\dagger} - \mathcal{L}_D\right)\]
where for the moment $\mathcal{L}_D^{\dagger}$ is defined formally and the factor of $-i$ is in line with the other natural first order operators $P_{\mu} =-i\partial_{\mu}$ and $\Pi_{\mu} = -i\nabla_{\mu}$ in the quantum mechanics.

Now $H$ is a second order differential operator, $D$ is first order and neither change fermion number, so the desired commutation relations
\[ \left[D,H\right] = 2iH, \quad \left[D,K\right]=-2iK, \quad \left[H,K\right]=-iD\]
force the special conformal generator $K$ to be zeroth order in derivatives and fermions, namely multiplication by some function $K(X)$.

Using the fact that $\left[\mathcal{L}_V,d\right]=0$ for any vector $V$, we find that
\begin{equation} \left[D,H\right] = 2iH \quad \Leftrightarrow \quad \left[\mathcal{L}_D,d^{\dagger}\right] = -2d^{\dagger},\label{eqn:LDddagger} \end{equation}
which can be reduced to a commutation relation with the Hodge star and hence to a statement about the flow of the volume form along $D$. The upshot is that we obtain the desired commutation relation if and only if the vector field $D$ is a homothety, so it satisfies
\begin{equation} \mathcal{L}_Dg = 2g \label{eqn:homothety}. \end{equation}
Unsurprisingly, since this didn't really rest on any details of supersymmetry, this is the same result as obtained in \cite{Michelson:Strominger:1999}.

For any vector field satisfying $\mathcal{L}_V g = \lambda g$ for constant $\lambda$ we can derive a much more explicit formula
\begin{equation} \left.\mathcal{L}_V^{\dagger}\right|_{\Omega^r} = (-1)^{r(m-r)+1}*\mathcal{L}_V* = \lambda(r-\tfrac{m}{2}) - \mathcal{L}_V\label{eqn:LVdagger}\end{equation}
for the adjoint. In particular, if $V$ is an isometry then $i\mathcal{L}_V$ is self-adjoint (this will be important later) and we find that
\[ \hat{D} = -i\mathcal{L}_D + i(r-\tfrac{m}{2}).\]
Using Cartan's formula for the Lie derivative
\[ \mathcal{L}_X = \left\{i_X,d\right\}\]
along with the quantisation dictionary (\ref{eqn:canonicaldictionary}) this can be written in QM language as 
\[ \hat{D} = \psi^{\mu \dagger}\Pi_{\mu}D^{\nu}\psi_{\nu} + D^{\mu}\psi_{\mu}\psi^{\nu \dagger}\Pi_{\nu} + i\left(\psi^{\mu \dagger}\psi_{\mu} - \tfrac{m}{2}\right).\]

The relation $\left[D,K\right] = -2iK$ is equivalent to\footnote{Of course, $K$ must be real so that $\hat{K}$ is hermitian.}
\begin{equation} \mathcal{L}_DK = 2K\label{eqn:LDK}\end{equation}
but the remaining relation $\left[H,K\right] = -iD$ is harder work. After a somewhat lengthy calculation, we find that for a form $\alpha$ this reduces to:
\[ \left[\Delta,K\right] \alpha = -\left\{d,i_{\widetilde{dK}}\right\}\alpha + \left\{d^{\dagger},dK\wedge\right\}\alpha = \left(\mathcal{L}_D^{\dagger} - \mathcal{L}_D\right)\alpha, \]
where $\widetilde{dK}$ is the vector field dual to $dK$, i.e $\widetilde{dK}^{\mu} = g^{\mu \nu}\partial_{\nu}K$. Using the Cartan formula and its adjoint, we find that this holds if and only if
\begin{equation} D_{\mu} = \partial_{\mu}K.\label{eqn:DdK}\end{equation}
Following \cite{Michelson:Strominger:1999} we call a homothety satisfying this extra constraint closed.\footnote{Exact might seem like a more sensible terminology, but there may be issues with $K$ being a globally well-defined function. See section \ref{subsection:Kaehlercase} for a discussion in the K\"{a}hler case.} Combining equations (\ref{eqn:LDK}) and (\ref{eqn:DdK}) we find that
\begin{equation} K = \tfrac{1}{2}g^{\mu \nu}\partial_{\mu}K\partial_{\nu}K = \tfrac{1}{2}||D|| ^2 \label{eqn:K}.\end{equation}

To sum up, we find the same conditions as \cite{Michelson:Strominger:1999}, namely that our model admits an $\mathfrak{so}(2,1)$ conformal invariance if and only if there is a closed homothety
\[ \mathcal{L}_Dg = 2g, \quad D_{\mu}=\partial_{\mu}K, \quad K=\tfrac{1}{2}||D||^2.\]

\subsection{Extension of $\mathcal{N}=(1,1)$ SUSY to $\mathfrak{su}(1,1|1)$ superconformal invariance} \label{subsection:Basiccase}
We now consider the superconformal version of the basic SUSY algebra on any Riemannian $M$. $\mathfrak{su}(1,1|1)$ is a simple real superalgebra with bosonic part
\[ \mathfrak{u}(1)\oplus \mathfrak{su}(1,1) \cong \mathfrak{u}(1)\oplus \mathfrak{so}(2,1)\]
and a total of 4 fermions, with charges $\pm 1/2$ under $J_3$, forming doublets $(Q,S)$ and $(Q^{\dagger},S^{\dagger})$ of the conformal algebra $(D,H,K)$. We have already seen that $Q$ and $Q^{\dagger}$ commute with $H$, and from (\ref{eqn:LDddagger}) we obtain
\[ \left[D,Q\right] = iQ, \quad \left[D,Q^{\dagger}\right] = iQ^{\dagger}.\]

We define the superconformal generators $(S,S^{\dagger})$ via the relation
\[ \left[K,Q\right] = iS\]
from which we immediately calculate
\begin{subequations} \label{eqn:superconformal} \begin{align}S &= idK\wedge = iD_{\mu}\psi^{\mu \dagger} \\ S^{\dagger} &= -ii_D = -iD^{\mu}\psi_{\mu}.  \end{align}\end{subequations}
It is straightforward to check consistency of the remaining commutators. The non-vanishing ones are\footnote{A complete list of commutation relations is given in the appendix.}
\begin{subequations} \begin{align} \left[ H, S\right] = -iQ, \quad \left[D, S\right] &= -iS, \quad
   \left[ J_3, Q\right] = \tfrac{1}{2}Q, \quad \left[J_3, S\right] = \tfrac{1}{2}S\\ \left\{ S, S^{\dagger}\right\} &= 2K, \quad \left\{Q, S^{\dagger} \right\} = D-2iJ_3
   \end{align} \end{subequations}
plus adjoints. In particular, notice that the $\left\{Q,S^{\dagger}\right\}$ relation is just a rewriting of the Cartan formula, so that this algebra really reflects the ``natural'' structure on $M$.

We find, in parallel with \cite{Michelson:Strominger:1999}, that any Riemannian manifold admitting a conformal symmetry in the sense of section \ref{subsection:conformal} also admits an $\mathcal{N}=(1,1)$ superconformal invariance $\mathfrak{su}(1,1|1)$ with a $U(1)$ R-symmetry acting only on fundamental fermions.

\subsection{The K\"{a}hler case: $\mathcal{N}=(2,2)$ SUSY to $\mathfrak{u}(1,1|2)$ SCQM} \label{subsection:Kaehlercase}
We now look to extend the construction of section \ref{subsection:Basiccase} to K\"{a}hler geometry. Here the relevant superalgebra turns out to be $\mathfrak{u}(1,1|2)$, which is not simple (more on this later). It has bosonic part
\[ \mathfrak{u}(1)\oplus\mathfrak{u}(1)\oplus \mathfrak{so}(2,1)\oplus \mathfrak{su}(2)\]
of which we will see that $SU(2)\times U(1)$ forms a genuine R-symmetry and the remaining $U(1)$ is a global symmetry. There are 8 fermions: the supercharges $(Q,Q^{\dagger},\tilde{Q},\tilde{Q}^{\dagger})$ and their $\mathfrak{so}(2,1)$ doublet partners, the superconformal generators.

We have two $\mathcal{N}=(1,1)$ subalgebras with common bosonic part: one as in section \ref{subsection:Basiccase} and one with all fermions replaced by their conjugation by $I$ as in (\ref{eqn:Iconjugation}). In particular, the second set of superconformal generators is
\begin{subequations} \label{eqn:superconformalI} \begin{align} \tilde{S} &= (\partial -\bar{\partial})K\wedge = -i\psi^{\mu \dagger}I_{\mu}^{~\nu}D_{\nu} = \psi^{\imath \dagger}D_{\imath} - \psi^{\bar{\imath} \dagger}\bar{D}_{\bar{\imath}} \label{eqn:SI}\\
 \tilde{S}^{\dagger} &= -i i_{D^I} = -iD^{\mu}I_{\mu}^{~\nu} \psi_{\nu} = D^{\imath}\psi_{\imath} - \bar{D}^{\bar{\imath}}\psi_{\bar{\imath}} \label{eqn:SdaggerI} \end{align}
\end{subequations}
where the indices $(\imath,\bar{\imath})$ correspond to a complex coordinate system $(Z^{\imath},\bar{Z}^{\bar{\imath}})$ and
\[\tilde{D}^{\mu} = I_{\nu}^{~\mu}D^{\nu}.\]
 
It is not entirely trivial that the second $\mathcal{N}=(1,1)$ subalgebra with conjugated fermions has the correct commutation relations. We can see either geometrically (since $d^I$ is the imaginary part of the Dolbeault operator $\partial$) or algebraically (from the Jacobi identity) that obtaining the relation
\[ \left[D,\tilde{Q}\right] = i\tilde{Q} \]
requires the homothety $D$ to respect the complex structure. More precisely, we find the equivalent conditions
\begin{equation} \left[D,R\right] = 0, \quad \mathcal{L}_DI = 0. \label{eqn:Dhol} \end{equation}
This says, as in \cite{Michelson:Strominger:1999}, that $D$ is a holomorphic vector field, which in complex coordinates is simply
\[ D = D^{\imath}(Z)\partial_{\imath} + \bar{D}^{\bar{\imath}}(\bar{Z})\bar{\partial}_{\bar{\imath}}.\]
This constraint, on top of those from section \ref{subsection:Basiccase}, turns out to be sufficient to give our second $\mathfrak{su}(1,1|1)$ subalgebra.

Notice that the constraint (\ref{eqn:Dhol}) implies a few more important geometric identities which we make heavy use of. For the homothety $D$ we have
\begin{equation} \mathcal{L}_Dg=2g, \quad \mathcal{L}_DI = 0 \quad \Rightarrow \quad \mathcal{L}_D\omega = 2\omega \label{eqn:LDomega} \end{equation}
while for $\tilde{D}$ we have
\begin{equation} \mathcal{L}_{\tilde{D}}I = \mathcal{L}_{\tilde{D}}\omega = \mathcal{L}_{\tilde{D}}g = \left[D,\tilde{D}\right] = 0, \label{eqn:DIisometry}\end{equation}
which says that $\tilde{D}$ is a holomorphic isometry. In particular, (\ref{eqn:LVdagger}) says that the operator
\begin{equation} \hat{\tilde{D}} = -i \mathcal{L}_{\tilde{D}} \label{eqn:DI} \end{equation}
is self-adjoint. Moreover, (\ref{eqn:DIisometry}) implies that $\tilde{D}$ commutes with absolutely everything and hence generates a $U(1)$ global symmetry.

We now have all the ingredients in place to construct the $\mathfrak{u}(1,1|2)$ superconformal algebra. It consists of the two $\mathfrak{su}(1,1|1)$ subalgebras described above together with the $\mathfrak{u}(1)$ global symmetry (\ref{eqn:DI}), the $\mathfrak{u}(1)$ R-symmetry (\ref{eqn:RI}) and the $\mathfrak{su}(2)$ R-symmetry (\ref{eqn:Kaehlersu2}). The bosonic parts of all these components commute, so it remains to check the cross-terms between the two sets of fermions.\footnote{See the appendix for a full list of commutation relations.}

As already explained in section \ref{section:SUSY}, the correct relations for the supercharges are guaranteed by the K\"{a}hler identities (\ref{eqn:Kaehleridentities}). The $\left\{S,S'\right\}$ relations are straightforward, so we have non-vanishing anticommutators
\begin{equation} \label{eqn:QQIandSSI} \left\{\tilde{Q},\tilde{Q}^{\dagger}\right\} = 2H, \quad \left\{\tilde{S},\tilde{S}^{\dagger}\right\} = 2K \end{equation}
The $\left\{Q,S'\right\}$ relations are more complicated. Consider in particular
\begin{equation} \left\{Q,\tilde{S}\right\}\alpha = \left[(\partial + \bar{\partial})(\partial - \bar{\partial})K\right] \wedge \alpha = -2\partial \bar{\partial}K \wedge \alpha \label{eqn:QSI} \end{equation}
for some r-form state $\alpha$. This must close into another element of the superalgebra, so we're left with two options, the first of which is to demand that $\partial \bar{\partial}K = 0$. This is an extremely strong constraint, forcing $K$ to be the real part of a holomorphic function $f$. But this in turn is incompatible with $dK = g(D,-)$, since $\partial_{\imath}K = \partial_{\imath}f$ is holomorphic while $D_{\imath} = g_{\imath \bar{\jmath}}D^{\bar{\jmath}}$ is most certainly not.

The second option is to ask that $\partial \bar{\partial}K$ is a $(1,1)$ form already appearing in our algebra. There is exactly one possible choice, namely the K\"{a}hler form $\omega = J_+$, so we deduce that $K$ is the K\"{a}hler potential
\begin{equation} i\partial \bar{\partial}K = \omega. \label{eqn:Kaehlerpotential} \end{equation}
Notice that this solution also generalises nicely to the hyper-K\"{a}hler case, where there is a potential which is K\"{a}hler with respect to each complex structure.\footnote{At least in our situation: this is discussed further in section \ref{subsection:hyperKaehler}.} The possibility of superconformal algebras corresponding to a hyper-K\"{a}hler potential was also explored as a torsion-free example in \cite{Michelson:Strominger:1999}.

The remaining commutators can be deduced from those already obtained by taking adjoints and splitting equations into holomorphic and antiholomorphic parts (which we are allowed to do since both $D$ and $\tilde{D}$ are holomorphic vector fields). The new nonzero relations are
\begin{subequations} \label{eqn:QSIrelations} \begin{align} \left\{Q,\tilde{S}\right\} &= -\left\{\tilde{Q},S\right\} = 2iJ_+ \\ \left\{Q,\tilde{S}^{\dagger}\right\} &= -\left\{\tilde{Q},S^{\dagger}\right\} = \tilde{D} \\ \left\{\tilde{Q},\tilde{S}^{\dagger}\right\} &= D -2iJ_3 \end{align} \end{subequations}
and adjoints.

This completes our construction of $\mathfrak{u}(1,1|2)$. In particular, we find that the $U(1)$ R-symmetry $R$ is not generated by anticommutators, and that the global symmetry $\tilde{D}$ is central, so the algebra is not simple. The relationship of this structure to the natural representation of $\mathfrak{u}(1,1|2)$ by supermatrices is given in the appendix.

To sum up, we find that our model admits a $\mathfrak{u}(1,1|2)$ superconformal invariance if and only if the target space is K\"{a}hler and the conditions for $\mathfrak{su}(1,1|1)$ invariance as in section \ref{subsection:Basiccase} are met, with the extra constraints that the homothety is holomorphic and $K$ is the K\"{a}hler potential.

As remarked earlier, the K\"{a}hler potential is not really a well-defined function and we should be careful what we mean by this. As a section of a line bundle over $M$, it is defined only up to addition of the real part of a holomorphic function $\phi(Z)$. Since our closed homothety only depends on the first derivative of $K$, such a transformation induces
\[ D^{\imath} \mapsto D^{\imath} + g^{\imath \bar{\jmath}}\partial_{\bar{\jmath}}\phi\]
which does not preserve the holomorphy of $D$ unless $\phi$ is constant. In our situation we can fix the problematic redundancy by asking that $K=-\mu$ is the moment map with respect to $\omega$ of our Hamiltonian isometry $\tilde{D}$, so
\[ dK = -i_{\tilde{D}}\omega.\]
In particular, such a $K$ is defined up to addition of a constant and the homothety it induces is holomorphic. This option will actually be forced upon us in the hyper-K\"{a}hler case.\footnote{Of course, defining moment maps globally requires a trivial cohomology group $H^1(M)$. We will ignore this subtlety in this paper, but it can be important.}

\subsection{The hyper-K\"{a}hler case: $\mathcal{N}=(4,4)$ SUSY to $\mathfrak{osp}(4|4)$ SCQM} \label{subsection:hyperKaehler}
Finally we come to the hyper-K\"{a}hler case. The superconformal algebra here is a real form of the simple algebra $\mathfrak{osp}(4|4)$ with bosonic part
\[ \mathfrak{so}(2,1)\oplus \mathfrak{su}(2)\oplus \mathfrak{usp}(4).\]
There are 16 fermions in total: the supercharges $Q^{i\alpha}$ and superconformal generators $S^{j\beta}$. As the notation suggests, these each transform in the irreducible real representation $\mathbf{4}\otimes \mathbf{2}$ of $\mathfrak{usp}(4)\oplus\mathfrak{su}(2)$, and each pair $(Q^{i\alpha},S^{i \alpha})$ forms an $\mathfrak{sl}(2,\mathbb{R})$ doublet. Slightly more explicitly, $Q^{i\alpha}$ contains the generators $(Q,Q^{\dagger},Q^a,Q^{a\dagger})$ where, as in section \ref{section:SUSY}, $a=1,2,3$ indexes the three complex structures $I^a$.

We have already constructed all the necessary ingredients of this algebra: we simply take three copies of the $\mathfrak{u}(1,1|2)$ K\"{a}hler case, one for each complex structure, with the additional requirement that the homothety $D$ is triholomorphic
\[\mathcal{L}_DI^a = 0.\]
The $\mathfrak{usp}(4) \cong \mathfrak{so}(5)$ R-symmetry is just the algebra (\ref{eqn:SUSYso5}) identified by Verbitsky \cite{Verbitsky:1990}.

Of course, there are cross-terms between the $\mathcal{N}=(2,2)$ subalgebras. In particular, it is no longer true that the $U(1)$ symmetries generated by the Killing vectors $D^a$ are central. They combine to form an $\mathfrak{su}(2)$, but this turns out to be the wrong $\mathfrak{su}(2)$ in that it doesn't decouple from $\mathfrak{usp}(4)$. We will see that this can be remedied by a smart choice of linear combinations of generators.

Some important geometric identities are:
\begin{subequations} \label{eqn:hKgeometry} \begin{align} i_{D^a}\omega^b =& -\delta^{ab}dK-\epsilon^{abc}d^cK \\ \mathcal{L}_{D^a}\omega^b =& -2\epsilon^{abc} \omega^c \\ \mathcal{L}_{D^a}I^b =& -2\epsilon^{abc}I^c \\ \left[D^a,D^b\right] =& -2\epsilon^{abc}D^c, \end{align} \end{subequations}
where $\omega^a$ is the K\"{a}hler form defined with respect to $I^a$.

The consistency of all three K\"{a}hler subalgebras requires the existence of a hyper-K\"{a}hler potential: a single function (at least up to the ambiguities discussed in section \ref{subsection:Kaehlercase}) $K$ which is a K\"{a}hler potential for each complex structure. In contrast to the K\"{a}hler case, this is no longer a trivial requirement. The relevant result is that of \cite{Hitchinetal:1987}, where it is shown that a hyper-K\"{a}hler potential exists if and only if there is an isometric $SU(2)$ action permuting the complex structures and generated by vector fields $X_I,X_J,X_K$ such that $X=IX_I$ is independent of choice of complex structure $I$. Furthermore, the hyper-K\"{a}hler potential is then the moment map with respect to this $SU(2)$ action, i.e
\[i_{X_I}\omega^I = i_{X_J}\omega^J = i_{X_K}\omega^K = -dK.\]
Our conditions (\ref{eqn:LDomega}, \ref{eqn:DIisometry}, \ref{eqn:hKgeometry}) are easily seen to imply this, taking $X = D$.

From the geometry (\ref{eqn:hKgeometry}) we immediately obtain
\begin{subequations} \begin{align} \left[\hat{D}^a,\hat{D}^b\right] &= 2i \epsilon^{abc}\hat{D}^c \label{eqn:Liesu2} \\ \left[D^a,J_+^b\right] &= 2i\epsilon^{abc}J_+^c \label{eqn:DaJ+} \\ \left[D^a,S^b\right] &= 2i\epsilon^{abc}S^c \label{eqn:DaSb} \end{align} \end{subequations} 
and adjoints. The remaining commutation relations can now all be obtained using the Jacobi identity and results from the K\"{a}hler case, and a complete list of the results is given in the appendix.

We still need to verify the decoupling of the $\mathfrak{su}(2)$ and $\mathfrak{usp}(4)$ subalgebras. Notice that the problematic commutators all have the form
\[ \left[D^a,\mathcal{O}^b\right] = \lambda \epsilon^{abc}\mathcal{O}^c = -\tfrac{1}{2}\left[ R^a,\mathcal{O}^b\right] \]
so that the $\mathfrak{su}(2)$ generated by
\begin{equation} T^a = D^a + 2R^a \label{eqn:Ta} \end{equation}
decouples from $\mathfrak{usp}(4)$. Furthermore, it is easy to verify that the fermions form the correct representations of each bosonic subalgebra. In terms of the objects $Q^{i\alpha}$ and $S^{j\beta}$, the fermion anticommutators can be expressed as:
\begin{subequations} \label{eqn:fermions} \begin{align}
\left\{Q^{i\alpha},Q_{j\beta}\right\} &= 2\delta^i_j\delta^{\alpha}_{\beta}H \\
\left\{S^{i\alpha},S_{j\beta}\right\} &= 2\delta^i_j\delta^{\alpha}_{\beta}K \\
\left\{Q^{i\alpha},S_{j\beta}\right\} &= \delta^i_j\delta^{\alpha}_{\beta}D -2i\mathcal{J}^i_{mj}t^m \delta^{\alpha}_{\beta} + j^{\alpha}_{a\beta}T^a\delta^i_j.
\end{align} \end{subequations}
Here $m=1,\dots,10$ labels the adjoint of $\mathfrak{usp}(4)$, for which $t^m$ are a basis of generators, and $j_a,\mathcal{J}_m$ are matrix generators of $\mathfrak{su}(2)$ and $\mathfrak{usp}(4)$ in the fundamental representation.

To conclude, we find that our model admits an $\mathfrak{osp}(4|4)$ superconformal invariance if and only if $M$ is hyper-K\"{a}hler and has a triholomorphic homothety $D$ whose dual 1-form is derived from the hyper-K\"{a}hler potential $K$. This structure reduces self-consistently to $\mathfrak{u}(1,1|2)$, the K\"{a}hler case, and to $\mathfrak{su}(1,1|1)$, the generic Riemannian case, by ``forgetting about the extra complex structures''.

\subsection{Example: instanton quantum mechanics} \label{subsection:example}
BFSS matrix mechanics \cite{Banks:Fischler:Shenker:Susskind:1996} describes M-theory in terms of the quantum mechanics of $k$ D0-branes, via the discrete light cone quantisation (DLCQ) technique. In \cite{Aharony:Berkooz:Seiberg:1997} this approach is applied to the $(2,0)$ superconformal field theory living on a stack of $N$ M5-branes. The DLCQ procedure leads to the worldvolume theory of $k$ D0-branes inside $N$ D4-branes, which at low energies reduces to quantum mechanics on the moduli space $\mathcal{M}_{k,N}$ of $k$ $SU(N)$ Yang-Mills instantons on $\mathbb{R}^4$ \cite{Douglas:1995}. The brane configuration preserves 8 supercharges and so the quantum mechanics is an $\mathcal{N}=(4,4)$ model of the type we are considering here. \cite{Aharony:Berkooz:Seiberg:1997} argues that in fact the DLCQ retains a superconformal invariance from the $(2,0)$ theory: we claim that it must fit into our construction since the supersymmetry is of the correct form and our expressions for $\hat{K}$ agree.

The moduli space of instantons is obtained as a hyper-K\"{a}hler quotient of flat space $\mathbb{R}^{4k(N+k)}$ via the ADHM construction \cite{Atiyah:Drinfeld:Hitchin:Manin:1978,Hitchinetal:1987}. Flat $4l$-dimensional space provides a trivial example of our construction since we can split the coordinates $X^{\mu}$ as $l$ quaternions $X^{i\alpha \dot{\alpha}}$ transforming in the $\mathbf{2}\otimes \mathbf{2}$ of $SU(2)$. The hyper-K\"{a}hler potential is just
\[ K = \tfrac{1}{2}\sum_{\mu}\left(X^{\mu}\right)^2\]
and the vector fields $D^a$ are induced from the right $SU(2)$ action. Explicitly
\[ D = X^{\mu}\partial_{\mu}, \quad D^a = -iX^{i\alpha \dot{\alpha}} \sigma^{a \dot{\beta}}_{\dot{\alpha}} \partial_{i\alpha \dot{\beta}}, \]
where $\sigma^a$ are the Pauli matrices.

The vector fields $D$ and $D^a$ preserve the moment map condition $\mu = 0$ for the auxiliary $U(k)$ action in the ADHM construction. Put another way, the ADHM equations, which in the notation of \cite{Dorey:Hollowood:Khoze:Mattis:2002} are the three $k \times k$ matrix equations
\[ \sigma^{c\dot{\beta}}_{\dot{\alpha}}\left(\bar{w}^{\dot{\alpha}}w_{\dot{\beta}} + \bar{a}'^{\dot{\alpha} \alpha}a'_{\alpha \dot{\beta}}\right)=\mu^c=0\]
with $a'_{\alpha \dot{\alpha}}$ an Hermitian $k \times k$ matrix and $w_{\dot{\alpha}}$ a complex $N\times k$ matrix , are homogeneous of degree 2 under uniform rescalings of the variables and invariant under right $SU(2)$ transformations. $D$ and $D^a$ therefore restrict naturally to vector fields on $\mathcal{M}_{k,N}$ via the projection
\begin{equation} \pi : \mu^{-1}(0) \rightarrow \frac{\mu^{-1}(0)}{U(k)}. \label{eqn:principalbundle} \end{equation}
As explained in \cite{Hitchinetal:1987}, the metric, K\"{a}hler forms and complex structures of $\mathcal{M}_{k,N}$ are also inherited from $\mathbb{R}^{4k(N+k)}$ in this way. We also know that, on the level set $\mu^{-1}(0)$, our homothety satisfies
\begin{equation} g(X_A,D) = \omega^a(X_A,D^a) = \mathcal{L}_{D^a}\mu^{Aa} = 0, \quad g(I^aX_A,D) = \mathcal{L}_{D}\mu^{Aa} = 2\mu^{Aa}=0,\label{eqn:Dorthogonality} \end{equation}
where $X_A$ are the vector fields generating the $U(k)$ action and $\mu^{Aa}$ is the moment map corresponding to $X_A$ and $\omega^a$. Note that this would not be true for a generic choice of level set $\mu^{-1}(\lambda)$, as one uses for the non-commutative resolution of $\mathcal{M}_{k,N}$, so we have a geometric understanding of the breaking of conformal invariance in this circumstance.

Since $D$ satisfies the orthogonality conditions (\ref{eqn:Dorthogonality}) its projection to the quotient space is essentially trivial, ensuring that the homothety and triholomorphy are preserved by the quotient construction. Similar statements can be made for the isometries $D^a$. It's important here that the K\"{a}hler structure on the quotient space is just defined by orthogonal projection from the principal $U(k)$-bundle (\ref{eqn:principalbundle}). Finally, we know that the hyper-K\"{a}hler potential $K'$ on the quotient space is just that of $\mathbb{R}^{4k(N+k)}$ evaluated on a solution of the ADHM equations  \cite{Hitchinetal:1987}. But it must also be the moment map for the $SU(2)$ action, so that
\[ dK' = -i_{D^a}\omega^a = g(D,-)\]
and the homothety remains closed.

These results are sufficient to guarantee the $\mathfrak{osp}(4|4)$ invariance of quantum mechanics on $\mathcal{M}_{k,N}$. Furthermore, since the K\"{a}hler potential is given in a simple way in terms of that of $\mathbb{R}^{4k(N+k)}$ it is possible, at least in principle, to give reasonably explicit formulae for the generators. In particular, our construction reproduces the expressions given in \cite{Aharony:Berkooz:Seiberg:1997}.

As a closing remark, notice that everything said above applies to more general hyper-K\"{a}hler quotients. Given an $\mathfrak{osp}(4|4)$ invariant system with a triholomorphic isometric action of a group $G$, we can take the hyper-K\"{a}hler quotient and obtain another one if the moment map equations $\mu = 0$ of the $G$-action are preserved by the homothety and its associated $SU(2)$ action.

\acknowledgments
We would like to thank Chris Blair, Alasdair Routh and Kenny Wong for helpful discussions, and are indebted to Nick Dorey for his advice throughout this project. We are supported by the STFC.

\begin{appendices}
Throughout this section $r$ is the degree of a differential form and, in the complex case, $(p,q)$ is its bidegree. $m$ is the real dimension of $M$ and $n$ is its complex dimension. All bosonic operators are self-adjoint. For the sake of brevity we only display non-vanishing commutators, and do not show commutators which are just given as adjoints of ones we have shown.
\section{$\mathfrak{su}(1,1|1)$}
This algebra occurs in the Riemannian case with $\mathcal{N}=(1,1)$ supersymmetry. The bosonic generators are:
\begin{subequations} \nonumber \begin{align} \hat{D} &= -i\mathcal{L}_D + i\left(p-\tfrac{m}{2}\right) \\&= D^{\mu}\psi_{\mu}\psi^{\nu \dagger}\Pi_{\nu} + \psi^{\mu \dagger}\Pi_{\mu}D^{\nu}\psi_{\nu} + i\left(\psi^{\mu \dagger} \psi_{\mu} -\tfrac{m}{2}\right) \\
H &= \tfrac{1}{2}\Delta = \tfrac{1}{2}\left(\psi^{\mu \dagger}\Pi_{\mu}\Pi_{\nu}\psi^{\nu} + \Pi_{\mu}\psi^{\mu}\psi^{\nu \dagger}\Pi_{\nu}\right) \\
\hat{K} &=  \tfrac{1}{2}||D||^2 = \tfrac{1}{2}g^{\mu \nu}\partial_{\mu}K\partial_{\nu}K \\
J_3 &= \tfrac{1}{2}\left(p-\tfrac{m}{2}\right) = \tfrac{1}{2}\left(\psi^{\mu \dagger}\psi_{\mu}-\tfrac{m}{2}\right).\\
\intertext{The fermionic generators are:}
Q &= d = i \psi^{\mu \dagger}\Pi_{\mu}, \quad Q^{\dagger} = d^{\dagger} = -i\Pi_{\mu}\psi^{\mu} \\
S &= idK\wedge = i\partial_{\mu}K\psi^{\mu \dagger}, \quad S^{\dagger} = -ii_D = -iD^{\mu}\psi_{\mu}.
\end{align} \end{subequations}
The non-vanishing commutators are:
\begin{subequations} \nonumber \begin{align}
&\left[D,H\right] = 2iH, \quad \left[D,K\right] = -2iK, \quad \left[H,K\right] = -iD \\
&\left[D,Q\right] = iQ, \quad \left[D,S\right] = -iS, \quad \left[H,S\right] = -iQ, \quad \left[K,Q\right] = iS\\
&\left[J_3,Q\right] = \tfrac{1}{2}Q, \quad \left[J_3,S\right] = \tfrac{1}{2}S\\
&\left\{Q,Q^{\dagger}\right\} = 2H, \quad \left\{S,S^{\dagger}\right\} = 2K, \quad \left\{Q,S^{\dagger}\right\} = D-2iJ_3.
\end{align} \end{subequations}

\section{$\mathfrak{u}(1,1|2)$} \label{appendix:Kaehler}
This algebra appears in the K\"{a}hler $\mathcal{N}=(2,2)$ case. As a real form of $\mathfrak{gl}(2|2)$ it is not simple, but has two $\mathfrak{u}(1)$ factors which could be quotiented out to produce the simple algebra $\mathfrak{psu}(1,1|2)$ \cite{Musson:2012}. In terms of the natural representation by supermatrices, the $\mathfrak{u}(1)$ factors are:
\begin{itemize}
\item The supertrace, which commutes with all bosonic generators but not with the fermions, hence corresponds to the $U(1)$ R-symmetry $R$. It is not generated by any commutators so can be self-consistently ignored, but there is no good reason to do so.
\item The trace, which is central so generates a $U(1)$ global symmetry corresponding to $\tilde{D}$. This is generated by anticommutators, so removing it requires a formal quotient. In particular, there is no way to construct a faithful representation of $\mathfrak{psu}(1,1|2)$ from an associative algebra, so in our case we are forced to include this $U(1)$. This discussion is analagous to what happens in $\mathcal{N}=4$ SYM in 4 dimensions.
\end{itemize}
In addition to those operators occurring in $\mathfrak{su}(1,1|1)$, we have bosonic operators: 
\begin{subequations} \nonumber \begin{align}
J_+ &= \omega \wedge = \tfrac{1}{2}\omega_{\mu \nu}\psi^{\mu \dagger}\psi^{\nu \dagger} \\
J_- &= \left(\omega \wedge\right)^{\dagger} = \tfrac{1}{2} \omega_{\mu \nu}\psi^{\nu}\psi^{\mu} \\
R &= \tfrac{1}{2}\left(p - q\right) = -\tfrac{i}{2}\psi^{\mu \dagger}I_{\mu}^{~\nu}\psi_{\nu} \\
\hat{\tilde{D}} &= -i\mathcal{L}_{\tilde{D}} = D^{\rho}I_{\rho}^{~\mu}\psi_{\mu}\psi^{\nu \dagger}\Pi_{\nu} + \psi^{\mu \dagger}\Pi_{\mu}D^{\rho}I_{\rho}^{~\nu}\psi_{\nu}.\\
\intertext{The new fermionic operators are:}
\tilde{Q} &= i(\bar{\partial}-\partial) = -i\psi^{\mu \dagger}I_{\mu}^{~\nu}\Pi_{\nu}, \quad
\tilde{Q}^{\dagger} = i(\partial^{\dagger}-\bar{\partial}^{\dagger}) = -i\Pi_{\nu}I^{~\nu}_{\mu}\psi^{\mu} \\
\tilde{S} &= (\partial - \bar{\partial})K \wedge = -i\partial_{\mu}KI^{~\mu}_{\nu}\psi^{\nu \dagger}, \quad
\tilde{S}^{\dagger} = -ii_{D^I} = -iD^{\mu}I_{\mu}^{~\nu}\psi_{\nu}.
\end{align} \end{subequations}
The new non-vanishing commutators are:
\begin{subequations} \nonumber \begin{align} &\left[J_+,J_-\right] = 2J_3, \quad \left[J_3,J_{\pm}\right]=\pm J_{\pm}, ~~~ \left[J_3,\tilde{Q}\right] = \tfrac{1}{2}\tilde{Q}, \quad \left[J_3,\tilde{S}\right] = \tfrac{1}{2}\tilde{S}\\
&\left[D,\tilde{Q}\right] = i\tilde{Q}, ~~~ \left[D,\tilde{S}\right] = -i\tilde{S}, ~~~ \left[H,\tilde{S}\right] = -i\tilde{Q}, \quad \left[K,\tilde{Q}\right] = i\tilde{S}\\
&\left[R,Q\right]=\tfrac{i}{2}\tilde{Q}, \quad \left[R,\tilde{Q}\right] = -\tfrac{i}{2}Q, \quad \left[R,S\right]=\tfrac{i}{2}\tilde{S}, \quad \left[R,\tilde{S}\right] = -\tfrac{i}{2}S\\
&\left[J_+,Q^{\dagger}\right] = \tilde{Q}, \quad \left[J_+,\tilde{Q}^{\dagger}\right] = -Q, \quad \left[J_+,S^{\dagger}\right] = \tilde{S}, \quad \left[J_+,\tilde{S}^{\dagger}\right] = -S\\
&\left\{\tilde{Q},\tilde{Q}^{\dagger}\right\} = 2H, \quad \left\{\tilde{S},\tilde{S}^{\dagger}\right\} = 2K\\
&\left\{Q,\tilde{S}^{\dagger}\right\} = -\left\{ \tilde{Q}, S^{\dagger}\right\} = \tilde{D}, \quad \left\{\tilde{Q},\tilde{S}^{\dagger}\right\} = D-2iJ_3\\
&\left\{Q,\tilde{S}\right\} = -\left\{\tilde{Q},S\right\} = 2iJ_+.
\end{align}
\end{subequations}

\section{$\mathfrak{osp}(4|4)$}
This simple algebra appears in the $\mathcal{N}=(4,4)$ hyper-K\"{a}hler case. More precisely, we get a real form with bosonic part $\mathfrak{sl}(2,\mathbb{R})\oplus \mathfrak{su}(2)\oplus \mathfrak{usp}(4)$. It contains three copies of the K\"{a}hler case whose operators and commutation relations can be read off from appendix \ref{appendix:Kaehler} by replacing each occurrence of the complex structure $I$ with $I^a$ $(a=1,2,3)$. We also define the linear combinations
\[ T^a = D^a + 2R^a\]
to decouple $\mathfrak{su}(2)$ from $\mathfrak{usp}(4)$.

The remaining non-vanishing commutators correspond to expressions with two different complex structures. These are:
\begin{subequations} \nonumber \begin{align}
&\left[T^a,T^b\right] = 2i\epsilon^{abc}T^c, \quad \left[J_+^a,J_-^b\right] = 2\left(\delta^{ab}J_3 -i\epsilon^{abc}R^c\right)\\
&\left[R^a,J_+^b\right] = -i\epsilon^{abc}J_+^c, \quad \left[R^a,R^b\right] = -i\epsilon^{abc}R^c\\
&\left[T^a,Q^b\right] = -i\left(\delta^{ab}Q - \epsilon^{abc}Q^c\right), \quad \left[T^a,S^b\right] = -i\left(\delta^{ab}S - \epsilon^{abc}S^c\right)\\
&\left[R^a,Q^b\right] = -\tfrac{i}{2}\left(\delta^{ab}Q +\epsilon^{abc}Q^c\right), \quad \left[R^a,S^b\right] = -\tfrac{i}{2}\left(\delta^{ab}S + \epsilon^{abc}S^c\right)\\
&\left[J_+^a,Q^{b\dagger}\right] = -\delta^{ab}Q - \epsilon^{abc}Q^c, \quad \left[J_+^a,S^{b\dagger}\right] = -\delta^{ab} - \epsilon^{abc}S^c\\
&\left\{Q^a,Q^{b\dagger}\right\} = 2\delta^{ab}H, \quad \left\{S^a,S^{b \dagger}\right\} = 2\delta^{ab}K, \quad
\left\{Q^a,S^b\right\} = -2i\epsilon^{abc}J_+^c\\
&\left\{Q^a,S^{b\dagger}\right\} =\delta^{ab}\left(D-2iJ_3\right)+\epsilon^{abc}\left(2R^c + T^c\right).
\end{align}\end{subequations}

\end{appendices}

\bibliography{SCQM}
\bibliographystyle{JHEP}
\end{document}